\documentstyle[11pt,newpasp,twoside,psfig,epsfig]{article}
\def\edcomment#1{\iffalse\marginpar{\raggedright\sl#1\/}\else\relax\fi}
\reversemarginpar

\begin{document}
\title{Are all young stars disc accretors?}
 \author{Jorick S. Vink$^1$, J. E. Drew$^1$, T. J. Harries$^2$, and R. D. Oudmaijer$^3$}
\affil{$^1$Imperial College, London, UK\\
      $^2$Exeter University, UK\\
      $^3$University of Leeds, UK}

\section{Introduction}
We now understand how low mass stars evolve through
cloud collapse and magnetically-channelled disc accretion, but whether
higher mass stars are also disc accretors is as yet unknown.
To answer this question, we need observations probing the 
near-star geometry of higher mass stars.

\section{Method of H$\alpha$ spectro-polarimetry} 
Until sub-milliarcsecond imaging becomes possible, 
H$\alpha$ spectro-polarimetry is the most powerful 
technique to explore the first few stellar radii around 
young stars.
The method is based on the expectation that the H$\alpha$ 
emission -- formed over a large volume in a circumstellar 
medium -- undergoes less electron scattering than the stellar 
continuum. 
As the line is then less polarized than the continuum, a change 
in the polarization across the line occurs -- we refer to this as the 
classical line-effect (analogous to that seen in classical Be stars).

\section{Results for the Herbig Ae/Be stars}

\subsection{A typical Herbig Be star showing a classical line-effect}
 
\includegraphics{mwc1080.ps}
\vspace{6.1cm}
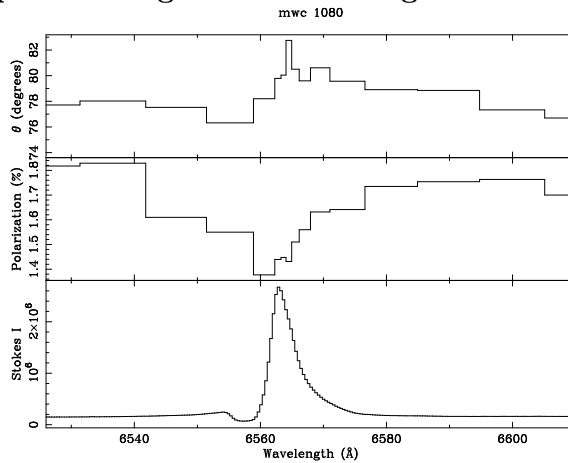
{\bf Figure 1:} Polarization spectrum (middle panel) of the Herbig Be star MWC 1080 (0.07\% binning). The position angle and Stokes I are also shown.

\noindent
Figure~1 shows the polarization spectrum (middle panel) 
for the typical Herbig Be star MWC 1080. 
Note the presence of the classical line-effect -- depolarization
across the line -- implying
that the electron-scattering region is not spherically symmetric,
but that the geometry is flattened, i.e. disc-like.

\subsection{A typical Herbig Ae star showing intrinsic line polarization}

\includegraphics{mwc480.ps}
\vspace{6.1cm}
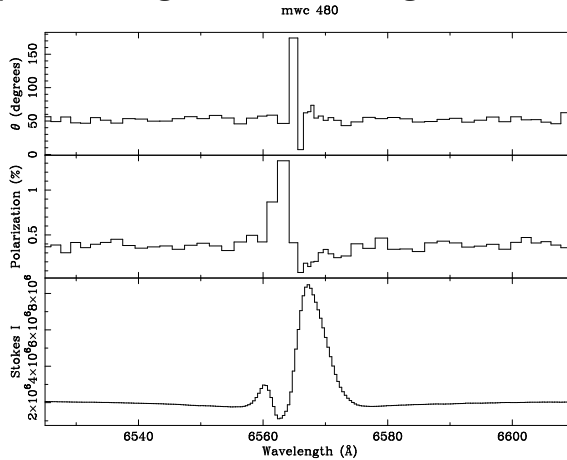
{\bf Figure 2:} Polarization spectrum of the Herbig Ae star MWC 480 (0.05\% binning).
The panels are as in Fig.~1. The polarization signature is more
complicated here, and is at variance with the simple depolarization 
picture described above.
The data for this and other Herbig Ae stars may suggest the presence of 
a compact H$\alpha$ source, located inside 
a rotating distribution of scatterers. 
Possibly, the magnetic accretor model -- commonly applied to 
low mass T~Tauri stars -- is able to explain the observations for 
the Herbig stars at this spectral type.

\subsection{Main results from our Herbig Ae/Be sample}
To be able to make a proper distinction between inclination and 
intrinsic geometrical effects, we have studied a larger sample (30+) 
of Herbig Ae/Be stars. So far, we 
find:
\begin{itemize}
\item For the Herbig Be stars: 10 out of 18 reveal depolarization, i.e. the classical line-effect. 
\item For the Herbig Ae stars: 9 out of 13 show an intrinsic line polarization effect.
\end{itemize}

\section{Conclusions}

Our data indicate that discs around Herbig Ae/Be stars are common, suggesting that the higher mass 
Herbig Ae/Be stars may well be disc accretors.
However, the Herbig Ae and the Herbig Be stars show different polarization signatures, 
possibly revealing differences in their modes of accretion.

\end{document}